# Surface plasmon on the metal nanosphere excitation by a quantum oscillator

A.M.Leontovich, V.V.Lidsky

The P.N.Lebedev Physical Institute of RAS
119991 Moscow, Russia
vlidsky@sci.lebedev.ru

Surface waves on a metal nanosphere are described in terms of quantum electrodynamics. The interaction of surface waves with a quantum oscillator is discussed in the dipole approximation. The increase in the spontaneous emission rate of the excited quantum oscillator, the so called Purcell factor, is evaluated to exceed in several orders that of in vacuum.

Расчет сферического плазмонного наноизлучателя
с учетом эффекта Перселла

А.М.Леонтович, В.В.Лидский

Физический институт им. П. Н. Лебедева РАН
119991 Москва, Ленинский пр., 53
vlidsky@sci.lebedev.ru

А Н Н О Т А Ц И Я

Рассмотрены поверхностные волны на идеальной сфере в рамках формализма квантовой электродинамики. Рассмотрено в дипольном приближении взаимодействие поверхностных плазмонов с квантовым излучателем, расположенным в непосредственной близости от поверхности сферы. Проведен расчет вероятности спонтанного перехода с излучением плазмона. Показано, что эта величина может на несколько порядков превосходить вероятность дипольного перехода в свободном пространстве.

1. После появления в печати сообщений о запуске нанолазеров или SPAZER'ов /1,2/, использующих механизм возбуждения поверхностных волн молекулами красителя, возник вопрос о построении последовательной квантовой теории взаимодействия поверхностных волн с квантовым излучателем. Дело в том, что вероятность спонтанного перехода квантовой системы существенным образом зависит от структуры пространства, окружающего излучатель. Перселл заметил, что вероятность спонтанного перехода возрастает на несколько порядков, если вблизи излучателя находится микроскопическая частица /3/. В /4/ эффект Перселла был привлечен для объяснения наблюдаемого увеличения на 14 порядков сечения комбинационного рассеяния — явления гигантского комбинационного рассеяния (SERS) /5,6/. Теория SPAZER'а была предложена в недавно появившейся работе /7/, где показано, что и здесь учет эффекта Перселла позволяет в принципе преодолеть трудность, вызванную исключительно сильным затуханием поверхностных волн.

Величина эффекта возрастания вероятности спонтанного излучения зависит от ряда факторов и в первую очередь сжатия волнового поля в поверхностных волнах, длина волны которых в десятки раз меньше длины волны соответствующей частоты в свободном пространстве. Поскольку концентрация энергии поля максимальна для сферических частиц, сравнительно с наносистемами низкой размерности, можно ожидать, что эффект Перселла для таких частиц окажется наиболее выраженным. Данная работа посвящена расчету эффекту Перселла для излучателя, расположенного вблизи наносферы.

Мы будем следовать методу расчета вероятности, описанному в [9]. Мы рассмотрим резонансные поверхностные волны на металлической сфере субволнового диаметра, окруженной диэлектрической средой. Затем мы разложим поле на элементарные моды и определим операторы вторичного квантования, с помощью которых поле поверхностной волны может быть описано с точки зрения квантовой электродинамики. Затем рассмотрим квантовый осциллятор, помещенный вблизи сферы и вычислим вероятность излучения этим осциллятором поверхностного плазмона. Полученную величину сравним с известной формулой для вероятности спонтанного излучения фотона в свободном пространстве и придем к формулировке эффекта Перселла, применительно к излучателю вблизи субволновой сферы.

2. Рассмотрим металлическую сферу малого радиуса $\rho$ в диэлектрической среде (рис. 1). Диэлектрическую проницаемость среды будем считать равной $\varepsilon_h$. Диэлектрическую проницаемость металла будем считать связанной с частотой формулой Друдэ: $\varepsilon_m = 1 - \omega_{pl}^2/\omega^2$. Мнимая часть диэлектрической проницаемости в данной работе учитываться не будет, так как учет поправок, связанных с затуханием в нашей постановке задачи оказался бы превышением точности. Магнитную проницаемость металла цилиндра будем считать равной 1.

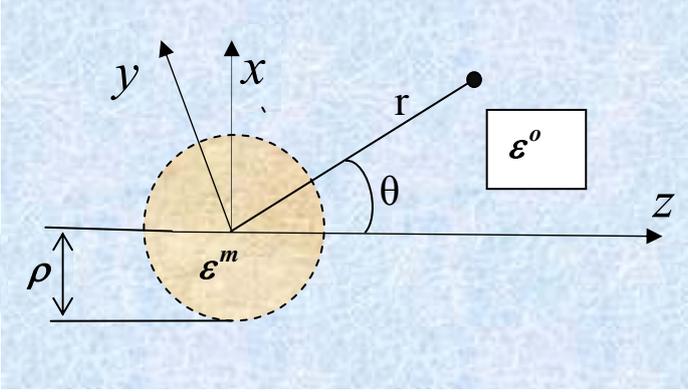

На первом шаге наша задача определить собственные поверхностные волны такого пространства. Уравнения Максвелла при отсутствии внешних токов запишем в виде:

$$rot\vec{H} = \varepsilon \cdot \partial_t \vec{E} \quad rot\vec{E} = -\partial_t \vec{H} \tag{2.1}$$

$$div\vec{H} = 0 \quad div\vec{E} = 0 \tag{2.2}$$

Напряженности $\vec{E}$ и $\vec{H}$, как известно, удовлетворяют волновому уравнению:

$$\Delta \vec{H} - \varepsilon \cdot \partial_t^2 \vec{H} = 0 \tag{2.3}$$

Рассмотрим монохроматическую волну, в которой зависимость напряженностей от времени определяется множителем $e^{-i\omega t}$:

$$\vec{E} = \vec{E}_\omega \cdot e^{-i\omega t} \quad \vec{H} = \vec{H}_\omega \cdot e^{-i\omega t} \tag{2.4}$$

Для упрощения записи мы будем опускать у компонент индекс $\omega$, там, где это не вызовет недоразумения. Волновое уравнение приходит к виду:

$$\Delta \vec{H} + \varepsilon \cdot \omega^2 \vec{H} = 0, \tag{2.5}$$

а его решениями оказываются собственные функции оператора Лапласа, который естественно представить в сферических координатах $(r, \vartheta, \varphi)$. Собственные функции угловой части зависят от двух индексов: $n = 0, 1, 2, ...$ и $-n \leq m \leq n$:

$$Y_{nm}(\vartheta, \varphi) = \frac{(-1)^m}{\sqrt{2\pi}} \cdot \sqrt{\frac{2n+1}{2} \cdot \frac{(n-m)!}{(n+m)!}} \cdot P_n^m(\cos\vartheta) \cdot e^{im\varphi}, \tag{2.6}$$

где $P_n^m(\xi)$ – присоединенная функция Лежандра, определенная для положительных $m$. Для вычисления $Y_{nm}(\vartheta, \varphi)$ при $m < 0$ будем использовать соотношение:

$$Y_{n,-m}(\vartheta, \varphi) = Y_{n,m}^*(\vartheta, \varphi) \tag{2.7}$$

Разложив компоненты напряженности по сферическим функциям, найдем уравнение, определяющее их зависимость от координаты $r$:

$$\frac{d^2}{dr^2}R_n + \frac{2}{r}\cdot\frac{d}{dr}R_n + \left(-\frac{n(n+1)}{r^2} + \varepsilon\omega^2\right)\cdot R_n = 0 \qquad (2.8)$$

Сделаем в (2.8) замену неизвестной функции:

$$R_n = r^{-\frac{1}{2}}\cdot B_n \qquad (2.9)$$

Функция $B_n$ удовлетворяет уравнению:

$$\frac{d^2}{dr^2}B_n + \frac{1}{r}\cdot\frac{d}{dr}B_n + \left(-\frac{(n+\frac{1}{2})^2}{r^2} + \varepsilon\omega^2\right)\cdot B_n = 0 \qquad (2.10)$$

В пространстве вне сферы ($r > \rho$) это уравнения имеет два независимых решения: функцию Бесселя $B_n = J_{n+\frac{1}{2}}(qr)$ и функцию Неймана $B_n = Y_{n+\frac{1}{2}}(qr)$ (не смешивать с обозначением сферической функции, имеющей два индекса!). Здесь через $q$ мы обозначили радиальную часть волнового вектора:

$$q = \omega\cdot\sqrt{\varepsilon^o} \qquad (2.11)$$

Внутри сферы ($\varepsilon < 0$) единственным регулярным решением (2.10) оказывается модифицированная функция Бесселя:

$$B_n = I_{n+\frac{1}{2}}(pr), \qquad (2.12)$$

где

$$p = \omega\cdot\sqrt{-\varepsilon^m}. \qquad (2.13)$$

Таким образом, вне сферы решение уравнения (2.5) есть линейная комбинация функций, которые мы обозначим как $\Omega_{nm}$:

$$\Omega_{nm}(r,\vartheta,\varphi) = r^{-\frac{1}{2}}\cdot B_{n+\frac{1}{2}}(qr)\cdot Y_{nm}(\vartheta,\varphi), \qquad (2.14)$$

где $B_\nu(z)$ – цилиндрическая функция, составленная из $J_\nu(z)$ и $Y_\nu(z)$. Решения уравнения (2.5) внутри сферы обозначим $\Theta_{nm}$:

$$\Theta_{nm}(r,\vartheta,\varphi) = r^{-\frac{1}{2}}\cdot I_{n+\frac{1}{2}}(pr)\cdot Y_{nm}(\vartheta,\varphi) \qquad (2.15)$$

3. Потенциалы поля выберем в «трехмерно-поперечной» калибровке, то есть потребуем, чтобы электрический потенциал $\varphi$ был равен нолю по всему пространству. Покажем, что это всегда возможно. Из уравнения (2.2) для $\vec{H}$ следует, что вектор $\vec{H}$ может быть представлен ротором некоторого вектора:

$$\vec{H} = rot\vec{A} \qquad (3.1)$$

Из (2.1) и (3.1) следует, что

$$rot(\vec{E} - i\omega\vec{A}) = 0 \qquad (3.2)$$

Это означает, что выражение в скобках есть градиент некоторой функции. Разделив этот градиент на $i\omega$ и прибавив к $\vec{A}$, что не изменит напряженностей $\vec{H}$ и $\vec{E}$, мы получим калибровку потенциала, в которой

$$\vec{E} = i\omega\cdot\vec{A} \qquad (3.3)$$

Откуда и вытекает равенство $\varphi = 0$. Из (3.3) также видно, что потенциал $\vec{A}$ в рассматриваемой калибровке удовлетворяет волновому уравнению (2.5).

Таким образом, мы приходим к следующему алгоритму решения волнового уравнения в пространстве со сферой. Выбираем в качестве потенциала $\vec{A}^o$ вне сферы произвольную комбинацию $\Omega_{nm}$, затем вычисляем ротор и определяем функции $\vec{H}^o$ вне сферы. Затем из (2.1) вычисляем $\vec{E}^o$ и с помощью найденного значения выполняем калибровку затравочного потенциала $\vec{A}^o$. Поскольку мы считаем материал частицы немагнитным ($\mu = 1$), на границе сферы непрерывны как тангенциальные, так и нормальная составляющая магнитного поля. Это позволяет легко вычислить компоненты напряженности магнитного поля внутри сферы:

$$\vec{H}^m_{nm} = \vec{H}^o_{nm} \cdot \frac{B_{n+\frac{1}{2}}(q \cdot \rho)}{I_{n+\frac{1}{2}}(p \cdot \rho)} \qquad (3.4)$$

Вычислив ротор выражения (3.4), воспользуемся (2.1) для определения компонент напряженности электрического поля $\vec{E}^m_{nm}$ внутри сферы. Для определения тангенциальных компонент $\vec{E}$ на границе сферы, вычислим векторное произведение $\left[ \vec{n}, \vec{E} \right]$, где $\vec{n}$ – вектор нормали к поверхности сферы:

$$\vec{n} = \{ \sin\vartheta \cdot \cos\varphi;\ \sin\vartheta \cdot \cos\varphi;\ \cos\vartheta \} \qquad (3.5)$$

Приравняем тангенциальные компоненты поля $\vec{E}^m_{nm}$ и $\vec{E}^o_{nm}$ на границе сферы. Характеристическое уравнение возникает как условие разрешимости полученной системы уравнений. Условие непрерывности нормальных компонент электрической и магнитной индукций следует из уравнений поля. Из характеристического уравнения может быть определено соотношение между диэлектрической проницаемостью металла на данной частоте и радиусом сферы.

Аналитический вид характеристического уравнения можно найти в [10], оно достаточно громоздко, мы не будем его выписывать. Мы реализовали указанный алгоритм на компьютере. Для вычисления производных по координатам мы воспользовались матричным представлением операторов дифферпенцирования в системах функций $\Omega_{nm}$ и $\Theta_{nm}$. Умножение на компоненты вектора нормали $\vec{n}$ также представляется соответсвующей матрицей. Отличные от ноля элементы этих матриц выписаны в Приложении А.

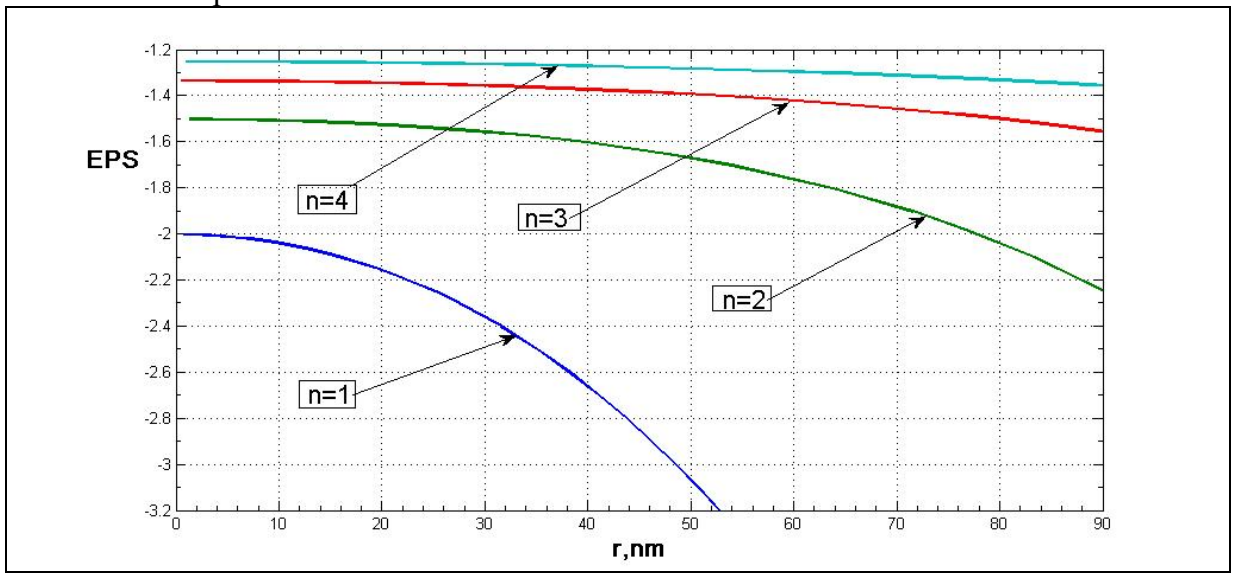

Рис. 1.
Характерристические кривые расчитанные для поверхностных волн различной мультипольности. По оси абсцисс отложен радиус наносферы в нанометрах в предоложении, что частота колебаний соответствует частоте света с длиной волны в вакууме 500 нм. По оси ординат – отношение диэлектрических проницаемостей металла и окружающего пространства: $EPS = \varepsilon^m / \varepsilon^o$

Полученные в результате расчета характеристические кривые представлены на рис. 1. Кривая с $n=1$ описывает колебания, где магнитное поле имет структуру дипольного типа, кривые с $n=2,3,4$ – описывают плазмоны соответствующей мультипольности. При интересующих нас малых радисах сферы, колебание с $n=1$ возникает при отношении диэлектрических проницаемостей близком к $-2$:

$$\frac{\varepsilon^m}{\varepsilon^o} \approx -2 \tag{3.6}$$

Предельное значение $\varepsilon^m$ для старших гармоник может быть вычислено по формуле:

$$\frac{\varepsilon^m}{\varepsilon^o} = -\frac{n+1}{n} \tag{3.7}$$

Для волн с четным номером гармоники $n$ имеется шесть независимых поляризаций, для волн с нечетным $n$ – три. В табл. 1. Приведены значения производящих потенциалов $\vec{A}$. Не вошедшие в табл. 1 независимые решения могут быть получены поворотом системы координат.

| $n=1$ | $A_{z,0,0} = 1$ | | | | | | |
|---|---|---|---|---|---|---|---|
| $n=2$ | $A_{x,1,0}=1;\ A_{y,1,0}=i;\ A_{z,1,1}=\sqrt{2}$ | | | | | | |
|  | $A_{x,1,0}=1;\ A_{y,1,0}=-i;\ A_{z,1,-1}=\sqrt{2}$ | | | | | | |
| $n=3$ | $A_{x,2,1}=1;\ A_{x,2,-1}=1;\ A_{y,2,1}=-i;\ A_{y,2,-1}=i;\ A_{z,0,0}=-\sqrt{\frac{6}{5}};$ | | | | | | |
| $n=4$ | $A_{x,1,0}=\frac{\sqrt{2}}{2};$ | $A_{y,1,0}=\frac{i\sqrt{2}}{2};$ | $A_{z,1,1}=1;$ | $A_{x,3,0}=\sqrt{\frac{7}{6}};$ | $A_{y,3,0}=i\sqrt{\frac{7}{6}};$ | $A_{x,3,2}=-\sqrt{\frac{7}{5}};$ | $A_{y,3,2}=i\sqrt{\frac{7}{5}}$ |
|  | $A_{x,1,0}=\frac{\sqrt{2}}{2};$ | $A_{y,1,0}=-\frac{i\sqrt{2}}{2};$ | $A_{z,1,-1}=1;$ | $A_{x,3,0}=\sqrt{\frac{7}{6}};$ | $A_{y,3,0}=-i\sqrt{\frac{7}{6}};$ | $A_{x,3,-2}=-\sqrt{\frac{7}{5}};$ | $A_{y,3,-2}=-i\sqrt{\frac{7}{5}}$ |

4. Пусть потенциал, определяющий поле вне сферы имеет вид:

$$\vec{A}^{(k)} = \gamma_k \vec{a}^{(k)} \tag{4.1}$$

Где $\vec{a}^{(k)}$ – в некоторый фиксированный вектор, определяющий поляризацию k-ой гармоники. Тогда энергия поля может быть представлена выражением:

$$W = \sum_k w^{(k)} \cdot \gamma_k \cdot \gamma_k^* \tag{4.2}$$

где $w^{(k)}$ — энергия волны, порождаемой вектором $\vec{a}^{(k)}$.

Величины $w^{(k)}$ вычислим, проинтегрировав по всему пространству плотность энергии электромагнитного поля:

$$w^{(k)} = \iiint_{r<\rho}\left(\frac{d(\omega\varepsilon^m)}{d\omega}\cdot\left(\vec{E}^{(k)}\right)^2 + \left(\vec{H}^{(k)}\right)^2\right)dV + \iiint_{r>\rho}\left(\varepsilon^o\cdot\left(\vec{E}^{(k)}\right)^2 + \left(\vec{H}^{(k)}\right)^2\right)dV \tag{4.3}$$

Первое слагаемое в (4.3) – энергия, заключенная в металлической частицы, для которой мы использовали формулу выражение Бриллюена для плотности энергии волны в среде с сильной дисперсией [10,§80] . Второе слагаемое соответствует энергии, сосредоточенной в диэлектрической среде.

Подставив в (4.3) разложение полей $\vec{E}^{(k)}$ и $\vec{H}^{(k)}$ по сферическим гармоникам и проинтегрировав по углам, с учетом ортогональности сферических функций, найдем:

$$w^{(k)} = \sum_{n=0}^{\infty} \sum_{m=-n}^{n} \left( \frac{d(\omega \varepsilon^m)}{d\omega} \cdot \vec{E}_{nm} \cdot \vec{E}_{nm}^* + \vec{H}_{nm} \cdot \vec{H}_{nm}^* \right) \cdot \int_0^{\rho} r \cdot \left( I_{n+\frac{1}{2}}(pr) \right)^2 dr$$
$$+ \sum_{n=0}^{\infty} \sum_{m=-n}^{n} \left( \varepsilon^o \cdot \vec{E}_{nm} \cdot \vec{E}_{nm}^* + \vec{H}_{nm} \cdot \vec{H}_{nm}^* \right) \cdot \int_{\rho}^{\infty} r \cdot \left( Y_{n+\frac{1}{2}}(qr) \right)^2 dr \quad (4.4)$$

Входящие в (4.4) интегралы могут быть вычислены по формулам:

$$\int_0^{\rho} r I_\nu^2(pr) dr = -\frac{\rho^2}{2} \cdot \left( I'_\nu(\rho p) \right)^2 + \frac{1}{2} \left( \rho^2 + \frac{\nu^2}{p^2} \right) \cdot I_\nu^2(p\rho) \quad (4.5)$$

$$\int_\rho^{\infty} r Y_\nu^2(qr) dr = -\frac{\rho^2}{2} \cdot \left( \left( Y_\nu(q\rho) \right)^2 - Y_{\nu-1}(q\rho) \cdot Y_{\nu+1}(q\rho) \right) \quad (4.6)$$

Для оценки производной диэлектрической проницаемости, входящей в (4.4) воспользуемся формулой Друдэ: $\varepsilon^m = 1 - \frac{\omega_{pl}^2}{\omega^2}$. Мы найдем

$$\frac{d(\omega \varepsilon^m)}{d\omega} = 2 - \varepsilon^m \quad (4.7)$$

На рис. 2 представлены результаты расчета величин $w^{(k)}$ в зависимости от радиуса сфера для первых четырех резонансных гармоник со значениями производящего потенциала из табл. 1. Мы видим, что величина энергии каждой гармоники быстро растет с уменьшением радиуса сферы и возрастает на несколько порядков при увеличении номера гармоники. Такое поведение объясняется характером роста функции Неймана $Y_{n+\frac{1}{2}}(qr)$.

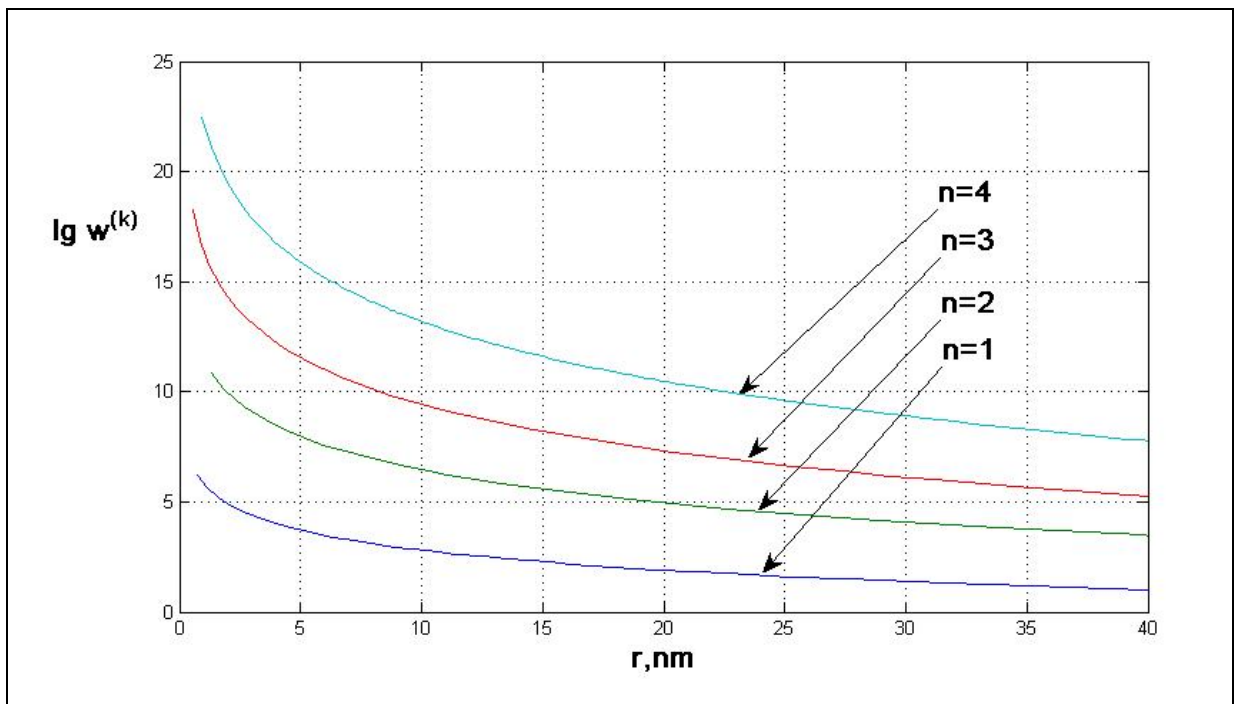

Рис. 2.
Значения полной энергии резонансного плазмона с производящими потенциалами $\vec{a}^{(k)}$ из табл. 1 в зависимости от радиуса сферы. Частота колебаний соответствует частоте света с длиной волны в вакууме 500 нм. По оси ординат отложен логарифм энергии в условных единицах.

5. Таким образом, система поверхностных волн на сфере описывается потенциалом:
$$\vec{A} = \sum_k \gamma_k \cdot a^{(k)} \cdot e^{-i\omega t} \tag{5.1}$$

Энергия таких волн выражается формулой (4.2). Наша задача описать систему поверхностных волн в рамках формализма квантовой электродинамики. Определим канонические переменные:
$$q_k = (\gamma_k \cdot e^{-i\omega t} + \gamma_k^* \cdot e^{i\omega t})\frac{\sqrt{2w^{(k)}}}{2\omega} \quad p_k = -i(\gamma_k \cdot e^{-i\omega t} - \gamma_k^* \cdot e^{i\omega t})\frac{\sqrt{2w^{(k)}}}{2} \tag{5.2}$$

Переменные $p_k, q_k$ удовлетворяют каноническим уравнениям:
$$\dot{q}_k = p_k; \quad \dot{p}_k = -\omega_k^2 q_k; \tag{5.3}$$

Функция Гамильтона системы поверхностных волн принимает вид:
$$H = \frac{1}{2}\sum_k \left(p_k^2 + \omega_k^2 \cdot q_k^2\right) \tag{5.4}$$

При переходе к квантовой теории мы должны рассматривать канонические переменные $p_k, q_k$ как операторы с правилом коммутации:
$$\widehat{p}_k\widehat{q}_k - \widehat{q}_k\widehat{p}_k = -i\hbar \tag{5.5}$$

Теперь мы можем выразить через операторы $\widehat{q}_k, \widehat{p}_k$ компоненты потенциала электромагнитного поля. Учитывая (5.1) и (5.2), находим:
$$\widehat{\vec{A}}^{(k)}(r,\vartheta,\varphi,t) = \frac{\vec{a}^{(k)}(r,\vartheta,\varphi)}{\sqrt{2w^{(k)}}} \cdot (\omega_k \widehat{q}_k + i\widehat{p}_k) \tag{5.6}$$

Гамильтониан системы плазмонов имеет вид вполне аналогичный (5.4):

$$\widehat{H} = \frac{1}{2}\sum_k \left(\widehat{p}_k^2 + \omega_k^2 \cdot \widehat{q}_k^2\right) \tag{5.7}$$

Гамильтониан распадается на сумму гармонических осцилляторов. Собственные значения такого гамильтониана хорошо известны:

$$E_n = \left(n + \frac{1}{2}\right) \cdot \hbar \omega_k \tag{5.8}$$

А для матричных элементов операторов $\widehat{q}_k, \widehat{p}_k$ переходов между собственными состояниями гамильтониана можно получить соотношения:

$$\langle n_k |\widehat{q}_k| n_k - 1\rangle = \langle n_k - 1 |\widehat{q}_k| n_k \rangle = \sqrt{\frac{\hbar}{2\omega_k} \cdot n_k}$$

$$\langle n_k |\widehat{p}_k| n_k - 1\rangle = -\langle n_k - 1 |\widehat{p}_k| n_k \rangle = i\sqrt{\frac{\hbar \omega_k}{2} \cdot n_k} \tag{5.9}$$

Вместо канонических операторов $\widehat{q}_k, \widehat{p}_k$ удобно использовать их линейные комбинации, имеющие только по одному отличному от ноля матричному элементу для переходов между собственными состояниями гамильтониана (5.7):

$$\widehat{c}_k = \frac{1}{\sqrt{2\hbar\omega_k}} \cdot (\omega_k \widehat{q}_k + i\widehat{p}_k); \quad \widehat{c}_k^+ = \frac{1}{\sqrt{2\hbar\omega_k}} \cdot (\omega_k \widehat{q}_k - i\widehat{p}_k); \tag{5.10}$$

Несложно показать, что

$$\langle n_k - 1 |\widehat{c}_k| n_k \rangle = \sqrt{n_k}; \quad \langle n_k |\widehat{c}_k^+| n_k - 1 \rangle = \sqrt{n_k}; \tag{5.11}$$

Правило коммутации для $\widehat{c}_k, \widehat{c}_k^+$ принимает вид:

$$\widehat{c}_k \widehat{c}_k^+ - \widehat{c}_k^+ \widehat{c}_k = 1 \tag{5.12}$$

Таким образом мы можем рассматривать операторы $\widehat{c}_k, \widehat{c}_k^+$ как операторы уничтожения и рождения плазмона.

Оператор вторично квантованного потенциала поля плазмона принимает вид:

$$\widehat{\vec{A}} = \sum_k \sqrt{\frac{\hbar \omega_k}{w^{(k)}}} \left(\widehat{c} \cdot \vec{a}^{(k)} \cdot \exp(-i\omega_k t) + \widehat{c}^+ \cdot \vec{a}^{(k)*} \cdot \exp(+i\omega_k t)\right) \tag{5.13}$$

6. Вычислим вероятность излучения плазмона в единицу времени излучателем, расположенным вблизи металлической сферы. Размеры излучателя будем считать малыми, сравнительно с длиной волны плазмона, и вероятность вычислим в дипольном приближении. Будем следовать методу расчета вероятности спонтанного излучения, изложенному в /9,§45/ для случая излучателя в свободном пространстве.

Взаимодействие электромагнитного поля с зарядом описывается оператором:

$$\widehat{V} = e \cdot \iiint dV \cdot \widehat{A}_\mu \widehat{j}^\mu \tag{6.1}$$

Здесь $e$ — заряд электрона, $\widehat{A}_\mu, \widehat{j}^\mu$ — вторично квантованные операторы потенциала электромагнитного поля и "плотности тока электрона". Плотность тока выражается через оператор волновой функции и матрицы Дирака:

$$\widehat{j}^\mu = \widehat{\overline{\psi}} \gamma^\mu \widehat{\psi} \tag{6.2}$$

Интегрирование в (6.1) выполняется по всему 3-пространству.

Согласно теории возмущений вероятность перехода в единицу времени системы между состояниями дискретного спектра выражается формулой:

$$w_{fi} = \frac{2\pi}{\hbar^2 \omega} |V_{fi}|^2 \tag{6.3}$$

Матричный элемент оператора возмущения $V_{fi}$ вычисляется с помощью невозмущенных волновых функций состояний *i* и *f*.

В дипольном приближении считают величины $A_\mu(x,y,z)$ медленно меняющимися в пределах характерных размеров излучателя и выносят их из-под знака интеграла (6.1). В результате матричный элемент оператора (6.1), соответствующий излучению плазмона, приобретает вид:

$$\langle n_k+1, f|\widehat{V}|n_k, i\rangle = -e \cdot \langle n_k+1|\widehat{\vec{A}}_\kappa(\vec{r}_{em})|n_k\rangle \cdot \iiint dV \cdot \psi_f^* \gamma^0 \vec{\gamma} \psi_i, \qquad (6.4)$$

где учтено, что потенциал выбран в «трехмерно-поперечной» калибровке. Через $\vec{r}_{em}$ обозначена точка, где расположен излучатель. С помощью (5.13) и (5.11) находим:

$$\langle n_k+1|\widehat{\vec{A}}_\kappa(\vec{r}_{em})|n_k\rangle = \sqrt{\frac{\hbar\omega_k}{w^{(k)}}} \cdot \vec{a}^{(k)*} \cdot \sqrt{n_k+1} \qquad (6.5)$$

Интеграл по объему в (6.4) в нерелятивистском приближении есть матричный элемент скорости электрона, вычисляемый по нерелятивистским собственным функциям. Матричный элемент скорости связан с матричным элементом координаты и частотой перехода: $\vec{v}_{fi} = -i\omega\vec{r}_{fi}$. Вводя дипольный момент системы $\vec{d} = e\vec{r}$, находим для квадрата модуля матричного элемента:

$$\left|\langle n_k+1, f|\widehat{V}|n_k, i\rangle\right|^2 = \left|\vec{d}_{fi} \cdot \vec{a}^{(k)*}(\vec{r}_{em})\right|^2 \cdot \frac{\hbar\omega_k^3}{w^{(k)}} \cdot (n_k+1) \qquad (6.6)$$

Учитывая симметрию сферы, примем среднее значение квадрата косинуса угла, входящего в скалярное произведение равным $\frac{1}{3}$. Тогда из (6.3) и (6.6) найдем:

$$w_{fi} = \frac{2\pi}{3\hbar}\left|\vec{d}_{fi}\right|^2 \cdot \left|\vec{a}^{(k)*}(\vec{r}_{em})\right|^2 \cdot \frac{\omega_k^2}{w^{(k)}} \cdot (n_k+1) \qquad (6.7)$$

Для вычисления полной вероятности перехода с испусканием плазмона необходимо просуммировать величины $w_{fi}$ по всем возможным конечным состояниям, что в данном случае сводится к умножению на $N_p$ – число различных поляризаций с данной энергией. Для вероятности спонтанного излучения ($n_k = 1$), находим:

$$w_{pl} = \frac{2\pi}{3\hbar}\left|\vec{d}_{fi}\right|^2 \cdot \left|\vec{a}^{(k)*}(\vec{r}_{em})\right|^2 \cdot \frac{\omega_k^2}{w^{(k)}} \cdot N_p \qquad (6.8)$$

Сравним полученное выражение с известной формулой вероятности дипольного излучения в свободном пространстве /8,§45/:

$$w_{ph} = \frac{4\omega^3}{3\hbar} \cdot \left|\vec{d}_{fi}\right|^2 \qquad (6.9)$$

Введем «фактор Перселла» $F$, характеризующий возрастание вероятности спонтанного излучения плазмона квантовым диполем, расположенным вблизи субволновой металлической сферы, относительно вероятности излучения фотона:

$$F_k = \frac{w_{pl}}{w_{ph}} = \frac{\pi}{2\omega_k} \cdot \frac{N_p}{w^{(k)}} \cdot \left|\vec{a}^{(k)*}(\vec{r}_{em})\right|^2 \qquad (6.10)$$

На рис. 3-4 представлены результаты расчета по формуле (6.10). На рис. 3 излучатель предполагается находящимся вблизи поверхности сферы. В этом случае вероятность перехода резко зависит от радиуса сферы, возрастая до величины $10^6$ для радиуса $r \approx 2 - 3 nm$, при этом слабо зависит от номера гармоники. Несколько иной характер кривой представлен на рис. 4, где излучатель предполагается отстоящим от поверхности сферы на расстоянии равном радиусу. Вероятность перехода с излучением плазмона первых двух гармоник на порядок выше, чем с излучением третьей и четвертой. Для гармоник 1 и 2 вероятность достигает величин $10^4$ при радиусе сферы $r \approx 2 - 3 nm$, снижаясь до $10^2$ при $r \approx 6 - 7 nm$.

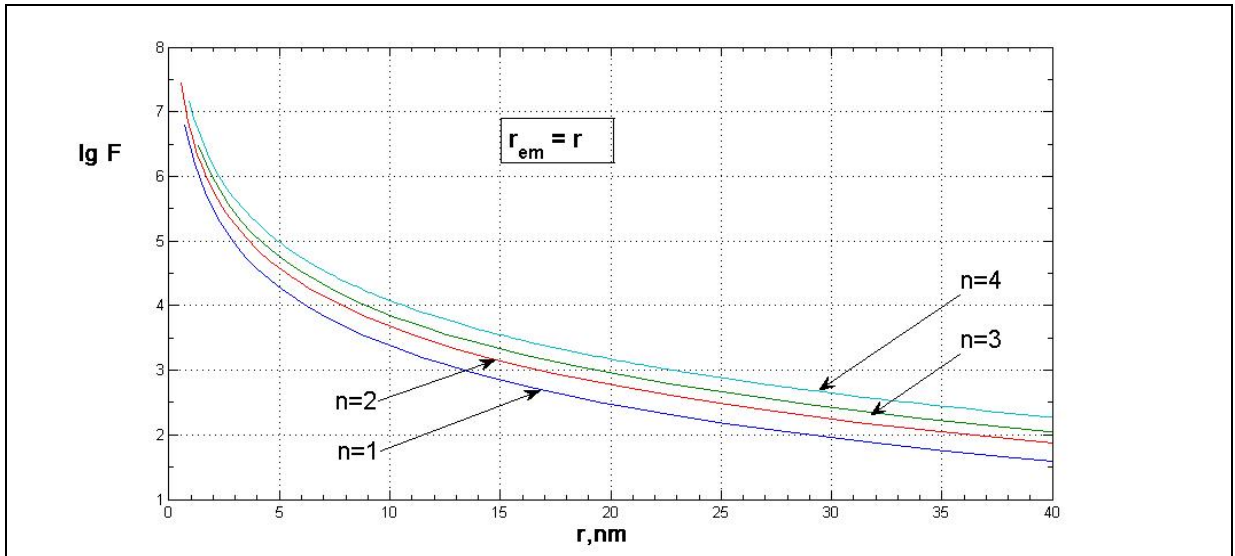

Рис.3

Фактор увеличения вероятности излучения резонансных плазмонов различной мультипольности квантовым осциллятором, расположенным вблизи поверхности наносферы. По оси абсцисс отложен радиус сферы.

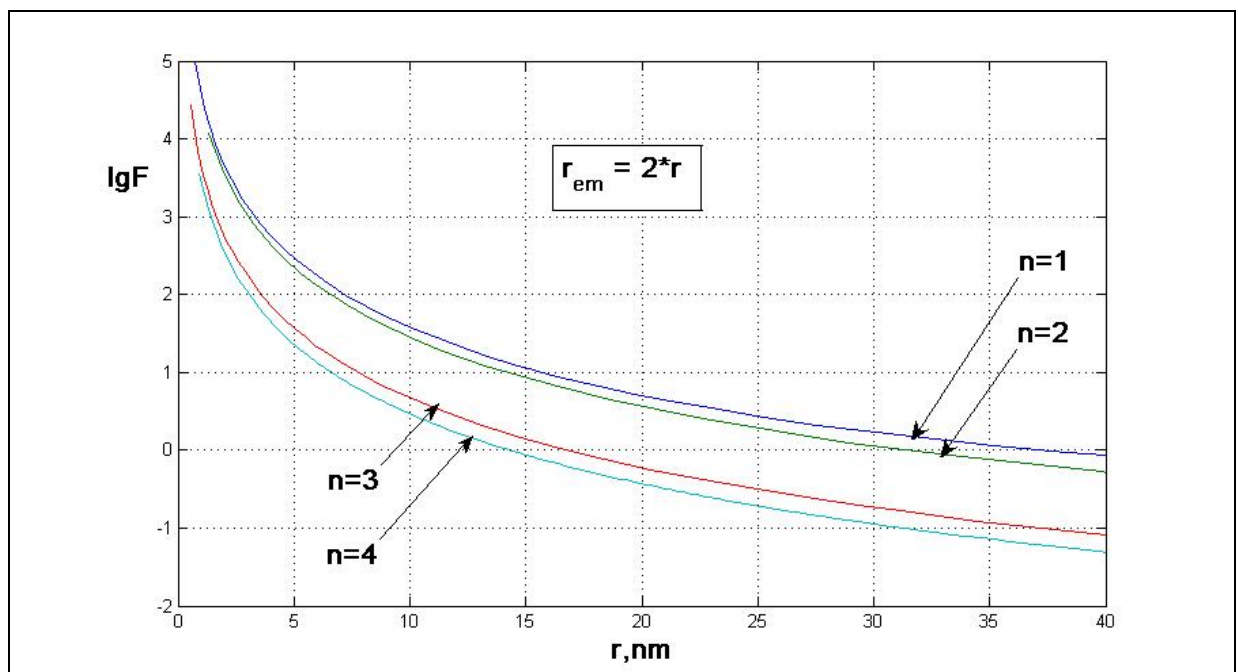

Рис.4
Фактор увеличения вероятности излучения резонансных плазмонов различной мультипольности квантовым осциллятором, расположенным на расстоянии радиуса сферы от ее поверхности. По оси абсцисс отложен радиус сферы.

6. Мы показали, что в присутствии сферических наночастиц увеличение вероятности спонтанного перехода излучающей частиц может достигать трех-шести порядков. Этот эффект необходимо учитывать при расчете наноразмерных лазеров или «спазеров».

Литература.

ПРИЛОЖЕНИЕ А.

Операторы дифференцирования по декартовым координатам $\{\partial_x, \partial_y, \partial_z\}$ в сферической системе координат принимают вид:

$$\partial_x = \sin\vartheta \cdot \cos\varphi \cdot \partial_r + \frac{1}{r} \cdot \cos\vartheta \cdot \cos\varphi \cdot \partial_\vartheta - \frac{1}{r \cdot \sin\vartheta} \cdot \sin\varphi \cdot \partial_\varphi$$

$$\partial_y = \sin\vartheta \cdot \sin\varphi \cdot \partial_r + \frac{1}{r} \cdot \cos\vartheta \cdot \sin\varphi \cdot \partial_\vartheta + \frac{1}{r \cdot \sin\vartheta} \cdot \cos\varphi \cdot \partial_\varphi \qquad (7.1)$$

$$\partial_z = \cos\vartheta \cdot \partial_r - \frac{1}{r} \cdot \sin\vartheta \cdot \partial_\vartheta$$

В системе функций $\Omega_{nm}$ (см. (2.14)) операторы дифференцирования представляются матрицами со следующими элементами:

$$\langle n-1, m-1 | \partial_x | n, m \rangle = \frac{q}{2} \cdot \sqrt{\frac{(n+m) \cdot (n+m-1)}{(2n-1) \cdot (2n+1)}}$$

$$\langle n-1, m+1 | \partial_x | n, m \rangle = -\frac{q}{2} \cdot \sqrt{\frac{(n-m) \cdot (n-m-1)}{(2n-1) \cdot (2n+1)}}$$

$$\langle n+1, m-1 | \partial_x | n, m \rangle = \frac{q}{2} \cdot \sqrt{\frac{(n-m+2) \cdot (n-m+1)}{(2n+3) \cdot (2n+1)}}$$

$$\langle n+1, m+1 | \partial_x | n, m \rangle = -\frac{q}{2} \cdot \sqrt{\frac{(n+m+2) \cdot (n+m+1)}{(2n+3) \cdot (2n+1)}}$$

$$\langle n-1, m-1 | \partial_y | n, m \rangle = \frac{iq}{2} \cdot \sqrt{\frac{(n+m) \cdot (n+m-1)}{(2n-1) \cdot (2n+1)}}$$

$$\langle n-1, m+1 | \partial_y | n, m \rangle = \frac{iq}{2} \cdot \sqrt{\frac{(n-m) \cdot (n-m-1)}{(2n-1) \cdot (2n+1)}}$$

$$\langle n+1, m-1 | \partial_y | n, m \rangle = \frac{iq}{2} \cdot \sqrt{\frac{(n-m+2) \cdot (n-m+1)}{(2 \cdot n+3) \cdot (2 \cdot n+1)}}$$

$$\langle n+1, m+1 | \partial_y | n, m \rangle = \frac{iq}{2} \cdot \sqrt{\frac{(n+m+2) \cdot (n+m+1)}{(2n+3) \cdot (2n+1)}}$$

$$\langle n-1, m | \partial_z | n, m \rangle = q \cdot \sqrt{\frac{(n+m) \cdot (n-m)}{(2n-1) \cdot (2n+1)}} \qquad (7.2)$$

$$\langle n+1, m | \partial_z | n, m \rangle = -q \cdot \sqrt{\frac{(n+m+1) \cdot (n-m+1)}{(2n+3) \cdot (2n+1)}}$$

В системе функций $\Theta_{nm}(r < \rho)$ (см. (2.15)) операторы дифференцирования представляются матрицами со следующими отличными от ноля элементами:

$$\langle n-1, m-1 | \partial_x | n, m \rangle = \frac{p}{2} \cdot \sqrt{\frac{(n+m) \cdot (n+m-1)}{(2n-1) \cdot (2n+1)}}$$

$$\langle n-1, m+1 | \partial_x | n, m \rangle = -\frac{p}{2} \cdot \sqrt{\frac{(n-m) \cdot (n-m-1)}{(2n-1) \cdot (2n+1)}}$$

$$\langle n+1, m-1 | \partial_x | n, m \rangle = -\frac{p}{2} \cdot \sqrt{\frac{(n-m+2) \cdot (n-m+1)}{(2n+3) \cdot (2n+1)}}$$

$$\langle n+1, m+1 | \partial_x | n, m \rangle = \frac{p}{2} \cdot \sqrt{\frac{(n+m+2) \cdot (n+m+1)}{(2n+3) \cdot (2n+1)}}$$

$$\langle n-1, m-1 | \partial_y | n, m \rangle = \frac{ip}{2} \cdot \sqrt{\frac{(n+m) \cdot (n+m-1)}{(2n-1) \cdot (2n+1)}}$$

$$\langle n-1, m+1 | \partial_y | n, m \rangle = \frac{ip}{2} \cdot \sqrt{\frac{(n-m) \cdot (n-m-1)}{(2n-1) \cdot (2n+1)}}$$

$$\langle n+1, m-1 | \partial_y | n, m \rangle = -\frac{ip}{2} \cdot \sqrt{\frac{(n-m+2) \cdot (n-m+1)}{(2 \cdot n+3) \cdot (2 \cdot n+1)}}$$

$$\langle n+1, m+1 | \partial_y | n, m \rangle = -\frac{iq}{2} \cdot \sqrt{\frac{(n+m+2) \cdot (n+m+1)}{(2n+3) \cdot (2n+1)}}$$

$$\langle n-1, m | \partial_z | n, m \rangle = p \cdot \sqrt{\frac{(n+m) \cdot (n-m)}{(2n-1) \cdot (2n+1)}} \tag{7.3}$$

$$\langle n+1, m | \partial_z | n, m \rangle = p \cdot \sqrt{\frac{(n+m+1) \cdot (n-m+1)}{(2n+3) \cdot (2n+1)}}$$

Небольшое различие в знаках в формулах (7.2) и (7.3) связано с различными рекуррентными формулами для функций $J_\nu(z), Y_\nu(z)$ и $I_\nu(z)$:

$$\begin{aligned} J_{\nu-1}(z) + J_{\nu+1}(z) &= \frac{2\nu}{z} J_\nu(z) \\ J_{\nu-1}(z) - J_{\nu+1}(z) &= 2J'_\nu(z) \\ Y_{\nu-1}(z) + Y_{\nu+1}(z) &= \frac{2\nu}{z} Y_\nu(z) \\ Y_{\nu-1}(z) - Y_{\nu+1}(z) &= 2Y'_\nu(z) \end{aligned} \tag{7.4}$$

$$\begin{aligned} I_{\nu-1}(z) - I_{\nu+1}(z) &= \frac{2\nu}{z} I_\nu(z) \\ I_{\nu-1}(z) + I_{\nu+1}(z) &= 2I'_\nu(z) \end{aligned} \tag{7.5}$$

Умножение на компоненты нормали (3.5) представляется следующими матрицами:

$$\langle n-1, m-1 | n_x | n, m \rangle = \frac{1}{2} \cdot \sqrt{\frac{(n+m) \cdot (n+m-1)}{(2n-1) \cdot (2n+1)}}$$

$$\langle n-1, m+1 | n_x | n, m \rangle = -\frac{1}{2} \cdot \sqrt{\frac{(n-m) \cdot (n-m-1)}{(2n+1) \cdot (2n-1)}}$$

$$\langle n+1, m-1 | n_x | n, m \rangle = -\frac{1}{2} \cdot \sqrt{\frac{(n-m+2) \cdot (n-m+1)}{(2n+1) \cdot (2n+3)}}$$

$$\langle n+1, m+1 | n_x | n, m \rangle = \frac{1}{2} \cdot \sqrt{\frac{(n+m+2) \cdot (n+m+1)}{(2n+1) \cdot (2n+3)}}$$

$$\langle n-1, m-1 | n_y | n, m \rangle = \frac{i}{2} \cdot \sqrt{\frac{(n+m) \cdot (n+m-1)}{(2n-1) \cdot (2n+1)}}$$

$$\langle n+1, m-1 | n_y | n, m \rangle = -\frac{i}{2} \cdot \sqrt{\frac{(n-m+2) \cdot (n-m+1)}{(2n+1) \cdot (2n+3)}}$$

$$\langle n-1, m+1 | n_y | n, m \rangle = \frac{i}{2} \cdot \sqrt{\frac{(n-m) \cdot (n-m-1)}{(2n+1) \cdot (2n-1)}}$$

$$\langle n+1, m+1 | n_y | n, m \rangle = -\frac{i}{2} \cdot \sqrt{\frac{(n+m+2) \cdot (n+m+1)}{(2n+1) \cdot (2n+3)}}$$

$$\langle n-1, m | n_z | n, m \rangle = \sqrt{\frac{(n-m) \cdot (n+m)}{(2n+1) \cdot (2n-1)}} \tag{7.6}$$

$$\langle n+1, m | n_z | n, m \rangle = \sqrt{\frac{(n-m+1) \cdot (n+m+1)}{(2n+1) \cdot (2n+3)}}$$

Поскольку умножение на компоненты $\vec{n}$ действует только на угловую часть собственной функции, выражения (7.6) применимы как с наружной, так и с внутренней стороны поверхности сферы.

Формулы (7.2), (7.3), (7.6) вытекают из рекуррентных соотношений (7.4)-(7.5) и следующих соотношений между присоединенными функциями Лежандра:

$$(2n+1) \cdot (1-z^2)^{\frac{1}{2}} \cdot P_n^m(z) = P_{n-1}^{m+1}(z) - P_{n+1}^{m+1}(z) \tag{7.7}$$

$$(2n+1) \cdot z \cdot P_n^m(z) = (n-m+1) \cdot P_{n+1}^m(z) + (n+m) \cdot P_{n-1}^m(z) \tag{7.8}$$

$$(2n+1) \cdot (1-z^2)^{\frac{1}{2}} \cdot P_n^{m+1}(z) = (n-m+1) \cdot (n-m) \cdot P_{n+1}^m(z) - (n+m+1) \cdot (n+m) \cdot P_{n-1}^m(z) \tag{7.9}$$

Формулы (7.7)-(7.9) доказываются непосредственно из определения присоединенной функции Лежандра:

$$P_n^m(\xi) = (-1)^m \cdot (1-\xi^2)^{\frac{m}{2}} \cdot \frac{d^m}{d\xi^m} P_n(\xi), \tag{7.10}$$

где $P_n(\xi)$ – полином Лежандра, который может быть вычислен по формуле Родрига:

$$P_n(\xi) = \frac{1}{2^n \cdot n!} \cdot \frac{d^n}{d\xi^n} \left(\xi^2 - 1\right)^n \tag{7.11}$$